\documentclass{aa}
\usepackage{graphicx}
\usepackage{txfonts}
\usepackage{natbib}
\bibpunct{(}{)}{;}{a}{}{,}
\begin{document}
   \title{Kelu-1\,AB -- A possible brown dwarf triple system}
   
   \subtitle{First resolved spectra and dynamical mass estimates}

   \author{M.B. Stumpf
          \inst{1},          
          W. Brandner\inst{1}, Th. Henning\inst{1}, H. Bouy\inst{2}, R. {K{\"o}hler}\inst{1,3}, F. Hormuth\inst{1,4}, V. Joergens\inst{1} \and M.Kasper\inst{5}
          }

   \offprints{M.B. Stumpf}

   \institute{Max-Planck-Institut f\"ur Astronomie, K\"onigstuhl 17,
              D-69117 Heidelberg, Germany\\
              \email{stumpf@mpia.de}
              \and
              Instituto de Astrofisica de Canaris, C/ V\'ia L\'actea s/n, E-38205 - La Laguna, Tenerife, Spain
              \and
              ZAH, Landessternwarte, K\"onigstuhl, D-69117 Heidelberg, Germany
              \and
              Centro Astron\'omico Hispano Alem\'an, C/Jes\'us Durb\'an Rem\'on 2-2$^\circ$, E-04004 Almer\'ia, Spain
              \and
              European Southern Observatory, Karl-Schwarzschild-Strasse 2, D-85748 Garching, Germany
          }

   \date{Received October 14, 2008 ; accepted --  }


  \abstract
   {Spatially resolved brown dwarf binaries provide the unique opportunity to determine the orbital motion and dynamical mass of the system independently of theoretical models. This is a key ingredient for the calibration of the mass-luminosity relations and evolutionary tracks.}
   {We have monitored the benchmark L dwarf binary Kelu-1\,AB over the past 3 years to derive improved spectral types and luminosities for the individual components. The astrometric measurements enable us to compute the orbital parameters and result in the first dynamical mass estimate for the system.}
   {We obtained resolved high angular resolution, near-IR images with HST and the VLT/NACO adaptive optics instrument in the J, H and K$_\textrm{s}$ bands. In addition we used NACO to achieve the first spatially resolved mid-resolution spectra in the H- and K-band for a precise spectral type determination.}
   {The near-IR spectrum of Kelu-1\,A reveals a distinct dip in the H-band providing evidence that Kelu-1\,A itself is a spectroscopic binary. We derive spectral types of L0.5 $\pm$ 0.5 for Kelu-1\,Aa and T7.5 $\pm$ 1 for Kelu-1\,Ab. Kelu-1\,B is classified as spectral type L3\,pec $\pm$1.5. From the relative orbit, we derive an orbital period of $38^{+8}_{-6}$ years and a semi-major axis of $6.4^{+2.4}_{-1.3}$ AU. This yields the first dynamically determined total system mass of $177^{+113}_{-55}\,M_{\rm Jup}$ for the Kelu-1\,AB system, with the uncertainties mainly attributed to the error of the system distance measurement and the yet missing apastron passage. The derived properties of Kelu-1\,AB allow us to test common theoretical models. The comparison of our results with color-magnitude diagrams based on evolutionary models yields a slightly revised age estimate (0.3 - 0.5 Gyr) and a discrepancy between dynamically and theoretically derived masses, stressing the importance for further dynamical mass determinations of brown dwarf binaries.}
   {}

\titlerunning{Kelu-1\,AB -- A possible brown dwarf triple system}
\authorrunning{M.B. Stumpf et al.}
   \maketitle
%

\section{Introduction}
Kelu-1 was discovered as one of the first single free-floating brown dwarfs (BDs) in the solar neighborhood by \citet{Ruiz97}. Originally classified with a spectral type L3 in the near-IR \citep{Knapp} at a distance of 18.7 $\pm$ 0.7\,pc and with a large proper motion of 285.0 $\pm$ 1.0 mas/yr \citep{Dahn}, it has been intensively studied since. More than 100 refereed publications listed in the SIMBAD database made it a kind of a benchmark object for L dwarfs, especially in spectral classification schemes. Particular interest was devoted to its apparent overluminosity compared to objects of similar spectral type \citep{Mart99_2, Leggett01, Goli04_2}, and to its photometric and spectroscopic variability with a 1.8\,h period in H$_{\alpha}$ \citep{Clarke02, Clarke03}. Since the presence of the \ion{Li}{i}\, absorption feature at 6708\,\AA\, confirmed its brown dwarf nature \citep{Ruiz97, Kirk99}, the brightness could only be explained by a surprisingly young age ($\sim$10 Myr, \citealt{Goli04_1}) or unresolved binarity, as first proposed by \citet{Mart99_2}.  Due to the free-floating nature of Kelu-1, its age is difficult to assess. In a first estimate \citet{Basri98_1} constrained the age on the basis of its \ion{Li}{i}\, absorption strength to 0.3-1 Gyr, which was later slightly revised by \citet{Liu} to 0.3-0.8 Gyr. This meant that a very young age was unlikely the cause of the overluminosity. However, \emph{Hubble Space Telescope} (HST) NICMOS observations of \citet*{Mart99_1} could not resolve any companion in a first attempt.

Six years later the mystery was solved. A binary companion was detected independently in March \citep{Gelino}  as well as in May 2005 \citep{Liu} with the Keck \emph{Laser Guide Star Adaptive Optics} (LGS AO) system. New (photometric) spectral types of L1.5-L3 for Kelu-1\,A and L3-L4.5 for Kelu-1\,B were determined, respectively. We were able to confirm the binary nature with our own HST/NICMOS data obtained in July 2005 and could already detect an orbital motion with an increase in separation by 0.015$\arcsec$ \,and in position angle (PA) by 0.43$\degr$ \,in five months. This made it a new promising L\,-dwarf binary target for dynamical mass determination, due to the first predicted reasonably short period of $\sim$\,35 years \citep{Liu}. 

Observational studies of brown dwarfs play a crucial role in the verification of evolutionary models and the theories of inner structure and atmospheres of substellar objects. As brown dwarfs never stabilize themselves on the hydrogen main sequence, there remains an ambiguity between the temperature or luminosity of a brown dwarf and its age or mass throughout its lifetime. However, spatially resolved binaries are expected to be coeval, thus removing part of the degeneracy in the mass-luminosity (age-temperature) relation which in turn can help to interpret their physical properties and test evolutionary models. In addition they provide the unique opportunity to determine the orbital motion and therefrom derive the dynamical mass of the binary and its components independently of any theoretical model. Although these dynamical mass measurements are strongly needed for very low mass stars (VLMS) and brown dwarfs, only few such measurements have been achieved up to now and large uncertainties remain (see \citealt{Zapa, Bouy04, Brand04, Stassun, Ireland, Liu08, Dupuy08}).

Therefore we started a detailed ground-based high-resolution adaptive optics photometric and spectroscopic observing campaign of the Kelu-1\,AB system, including a long-term monitoring program to measure the orbital motion and derive a meaningful determination of the orbital parameters. 
In this paper we present the first dynamical mass estimate for the binary system Kelu-1\,AB independent of theoretical models. In addition we show the first resolved spectra for each component, leading to improved spectral types and evidence for a possible unresolved third component in the system.


\section{Observations and data reduction}
The results of this paper are based on a variety of observations with different telescopes. For infrared photometry and astrometry we used our own HST/NICMOS and \emph{Very Large Telescope} (VLT) NACO observations as well as archived SPITZER observations. The infrared spectroscopy was carried out by means of own NACO data. Each observation will be consecutively described.

\subsection{HST/NICMOS Photometry}
\label{PhotHST}
KELU-1\,AB was observed on July 31, 2005 as part of our HST spectral differential imaging program GO 10208 (PI: W. Brandner), targeting 12 isolated L dwarfs with no known companions so far. The sources were observed with the NICMOS1 (NIC1) camera, providing a high-resolution pixel scale of 0\,\farcs0432 and a field of view (FoV) of 11\arcsec \,x 11\arcsec, in the two narrowband filters F108N (1.07 $\mu$m) and F113N (1.12 $\mu$m). We received 4 different images for the object in each filter. Two images were obtained in the same orbit, but at different positions on the detector, to be able to optimize PSF sampling, reject bad pixel and gain redundancy against cosmic ray events. All data were acquired in MULTIACCUM mode and the total integration times per filters were 2560\,s (F108N) and 2816\,s (F113N), respectively. 
   
The HST data analysis of the resolved binary is based on pipeline reduced frames as provided by the HST archive. To derive the magnitudes from the HST/NIC1 images we used aperture photometry with the IRAF \emph{phot} routine in the \emph{daophot} package. With a separation of 0.3\arcsec\, the distance between our components was less than the defined 0.5\arcsec\, (\,=\,11.5 pixels) separation for well-isolated sources. Thus we used an aperture size of 3 pixels and corrected to 11.5 pixel photometry using an aperture scaling factor derived from TinyTim\footnote{http://www.stsci.edu/software/tinytim/tinytim.html} PSF measurements. Then we adjusted our results to a nominal infinite aperture and converted the count rates into flux using the most recent photometric keyword-value as provided by the STScI webpage\footnote{http://www.stsci.edu/hst/nicmos/performance/photometry/postncs\_keywords.html}. For better comparison, we finally converted the flux into the Vega magnitude scale, using the flux zero points of 1937.0 and 1820.9 Jy for the F108N and F113N filters, respectively.  

We also reexamined the first HST/NICMOS observation of Kelu-1 from 14 August 1998 (GO 7952, PI E.Mart\'in). We retrieved 2 images from the HST archive which were acquired in MULTIACCUM mode with a total integration time of 448s each.

\subsection{VLT/NACO}
\subsubsection{Photometry}
Follow-up imaging observations were obtained with the adaptive optics system NACO at the ESO VLT/UT4 (Yepun) in the \emph{J} (1.25 $\mu$m), \emph{H} (1.65 $\mu$m) and \emph{K$_\mathrm{S}$} (2.15 $\mu$m) broad-band filters on April 28, 2006 as well as in L (3.8 $\mu$m) and NB4.05 (4.051 $\mu$m) band filters on May 26, 2006. Between May 2006 and June 2008 we acquired 6 additional epochs of K$_\mathrm{S}$ broad-band images within the framework of our ongoing astrometric monitoring program, as listed in Table\,\ref{ObsLog}. 

\begin{table}
\caption{Observation log of high-angular resolution imaging of the resolved binary Kelu-1\,AB}
\label{ObsLog}
\centering
\begin{tabular}{c c c c }     
\hline\hline
\noalign{\smallskip}
Date & Telescope/Instrument & Filter & Exp. time\\
 & & & [sec] \\
\hline
\noalign{\smallskip}
   31\,/\,07\,/\,2005 & HST/NIC1 & F108N & 2560 \\
   & & F113N & 2816  \\
   28\,/\,04\,/\,2006 & VLT/NACO& J & 8 x 80  \\
    & & H   &  8 x 60    \\
    & & K$_\mathrm{S}$   & 6 x 40   \\
   26\,/\,05\,/\,2006 & VLT/NACO & L\arcmin & 8 x 43.75  \\
   & & NB4.05 & 8 x 60 \\
   12\,/\,08\,/\,2006 & VLT/NACO & K$_\mathrm{S}$ & 8 x 55  \\
   22\,/\,02\,/\,2007 & VLT/NACO & K$_\mathrm{S}$ & 8 x 22 \\
   15\,/\,05\,/\,2007 & VLT/NACO & K$_\mathrm{S}$ & 8 x 30 \\
   13\,/\,03\,/\,2008 & VLT/NACO & K$_\mathrm{S}$ & 8 x 30 \\
   09\,/\,05\,/\,2008 & VLT/NACO & K$_\mathrm{S}$ & 8 x 20 \\
   03\,/\,06\,/\,2008 & VLT/NACO & K$_\mathrm{S}$ & 8 x 30 \\
\noalign{\smallskip}
\hline
\end{tabular}
\end{table}

We used the CONICA S27 (for \emph{J,H,K$_{S}$}) and L27 (\emph{L, NB4.05}) camera, providing a FoV of 28\arcsec\,x 28\arcsec \,with a pixel scale of 0\,\farcs0271. Wavefront sensing was performed on the primary component in the near-IR using the N90C10 dichroic for the observations in the \emph{J,H,K$_{S}$} bands and the JHK dichroic for the \emph{L, NB4.05} band observations. For sky computation and to optimize bad pixel/cosmic ray rejection the observations were executed in a jitter pattern with 6-8 points (see Table\,\ref{ObsLog}). To perform accurate photometry and astrometry, we chose the nearby point source 2MASS J13054388-25403805 from the GSC-II as PSF calibrator, which was observed in the same nights with the same instrumental settings as the binary.

  \begin{figure*}
  \centering
    \setlength{\unitlength}{1cm}
    \begin{minipage}[t]{3.4cm}
       \includegraphics[width=\textwidth]{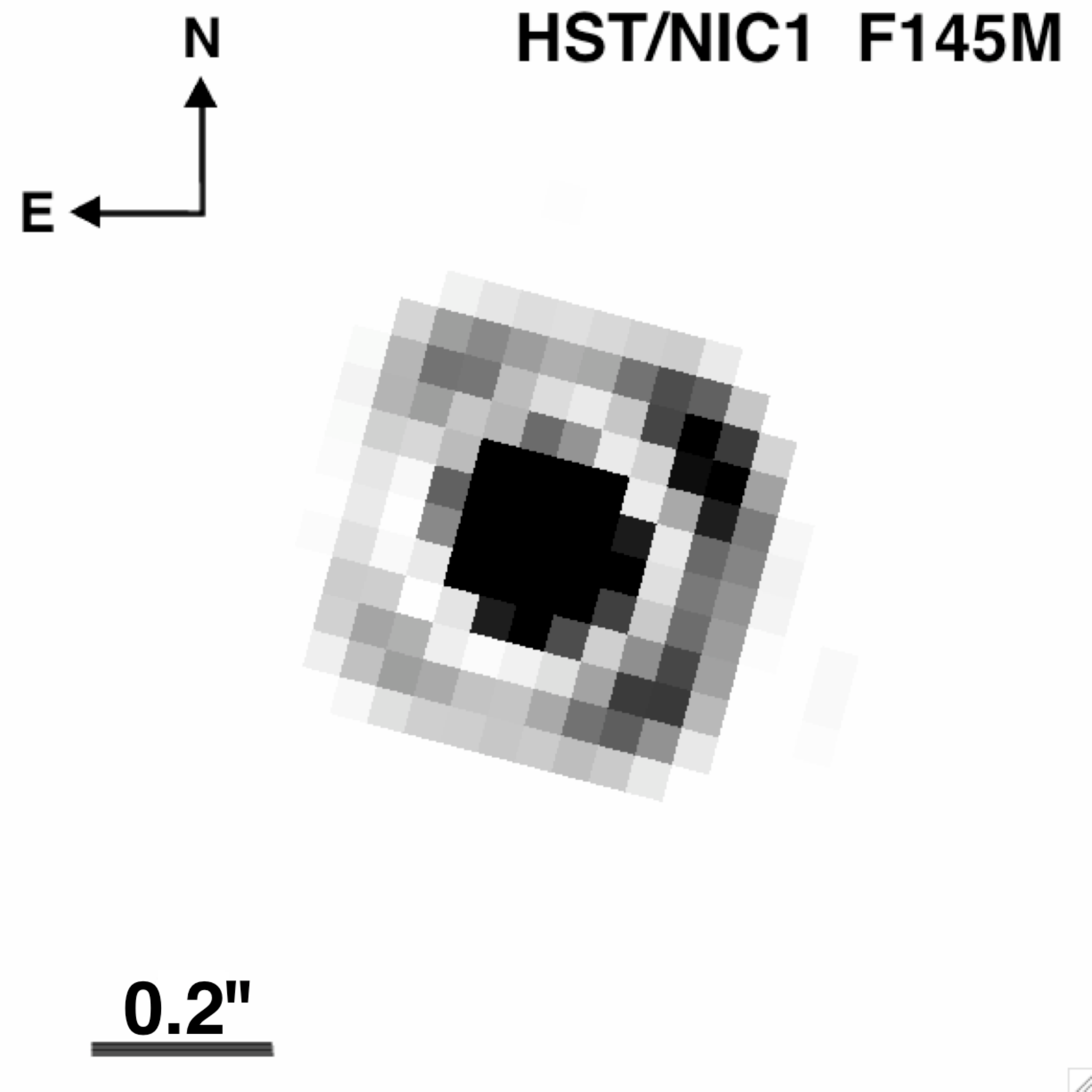}
       \begin{center}
         14\,/\,08\,/\,1998
       \end{center}
     \end{minipage}
    \begin{minipage}[t]{3.4cm} 
       \includegraphics[width=\textwidth]{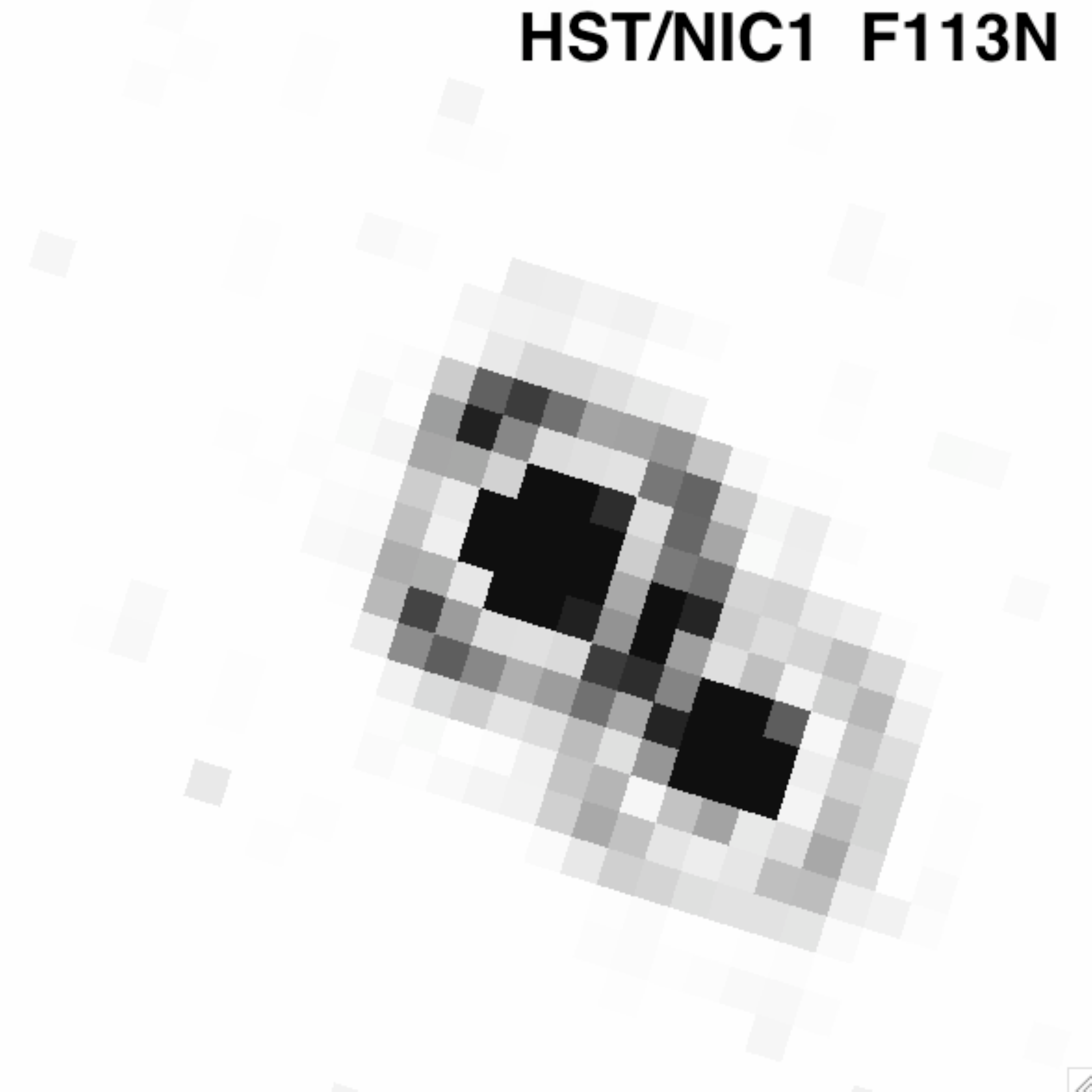}
       \begin{center}
         31\,/\,07\,/\,2005
       \end{center}
     \end{minipage}
      \begin{minipage}[t]{3.4cm} 
       \includegraphics[width=\textwidth]{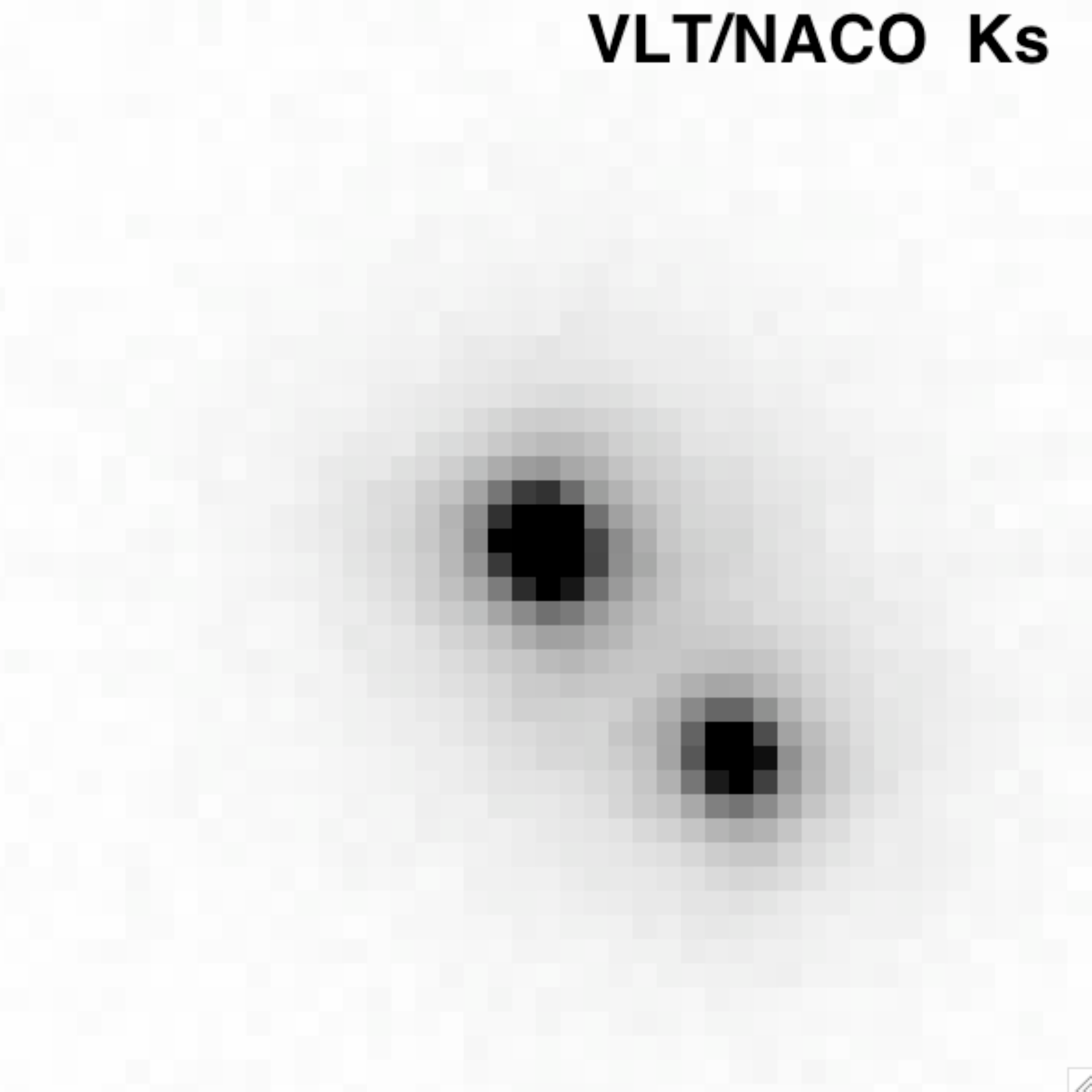}
        \begin{center}
         28\,/\,04\,/\,2006
       \end{center}
     \end{minipage}
      \begin{minipage}[t]{3.4cm} 
       \includegraphics[width=\textwidth]{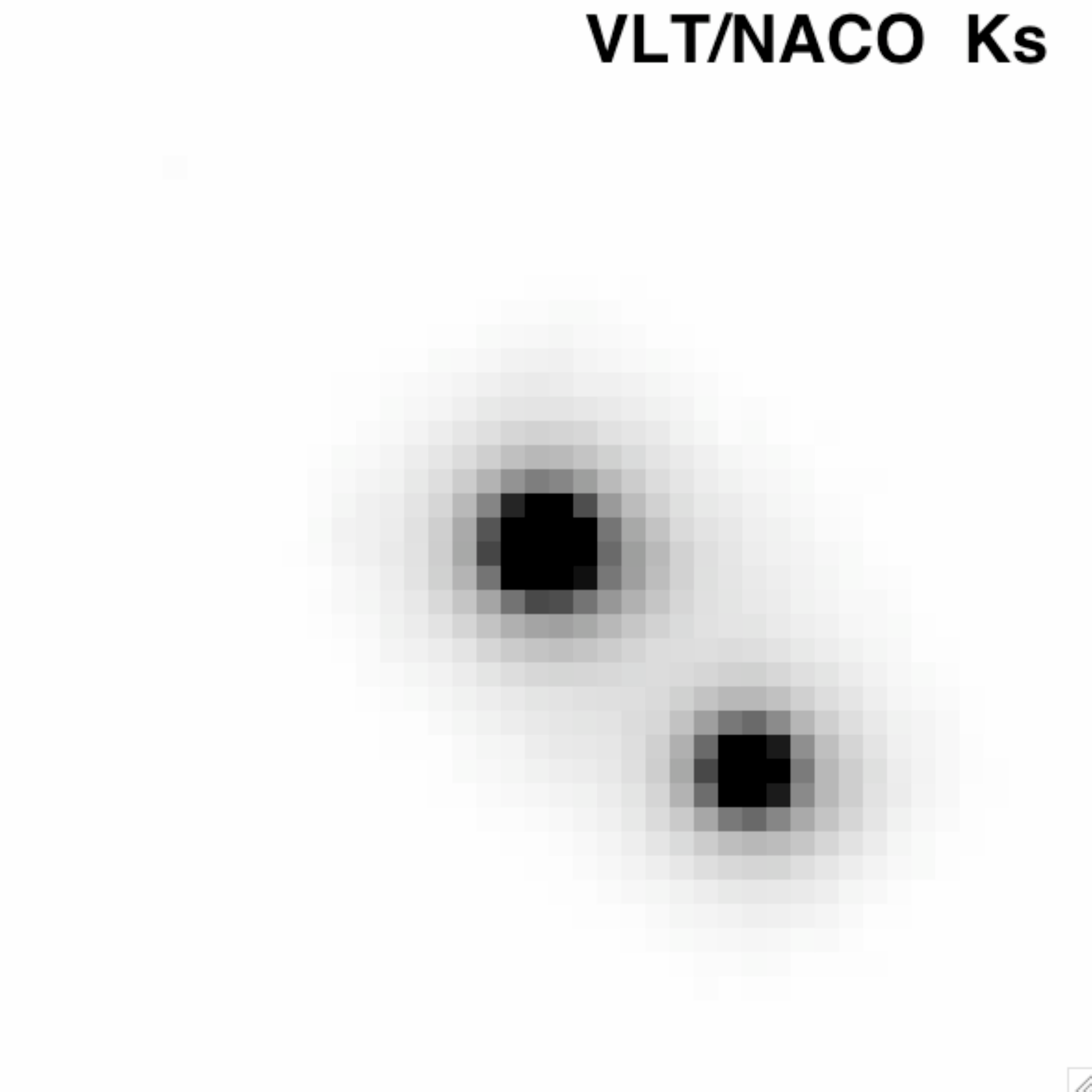}
        \begin{center}
         22\,/\,02\,/\,2007
       \end{center}
     \end{minipage}
    \begin{minipage}[t]{3.4cm} 
       \includegraphics[width=\textwidth]{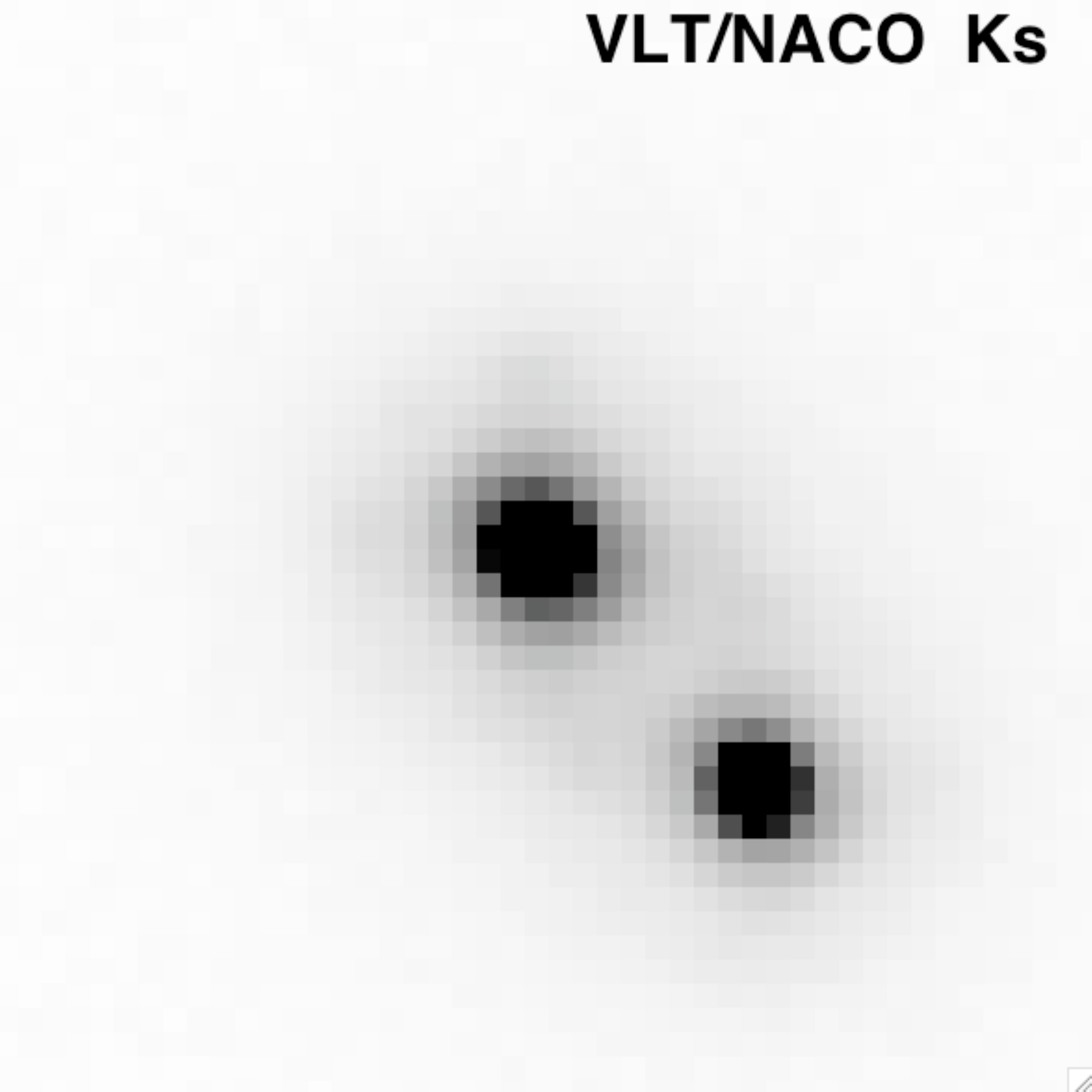}
        \begin{center}
         18\,/\,03\,/\,2008
       \end{center}
     \end{minipage}
       \caption{Images of Kelu-1\,AB obtained with HST/NICMOS and VLT/NACO at different epochs. The orientation and scale are the same for all images. Kelu-1\,A is the northern component while Kelu-1\,B is the component to the south west.}
   \label{ObsImages}
  \end{figure*}
%
On average the observations were executed under good atmospheric conditions with a very good seeing of $\sim$ 0.6$\arcsec$ FWHM leading to a Strehl ratio typically between 25-35 \% in K$_\mathrm{S}$. Only the observation on August 12, 2006 had slightly worse conditions with a seeing of 0.8$\arcsec$ FWHM and a Strehl ratio of only 15\%, resulting in a larger error estimation for the astrometry.

For the NACO/VLT observations the standard image processing including flat-fielding, bad pixel cleaning, dark and sky correction as well as average combination, was accomplished using the recommended Eclipse \emph{jitter} \citep{Devil} software package. We obtained very good final images in almost all filters with the exception of the images in the NB4.05 filter, where we could only barely detect the Kelu-1\,AB system due to a very low S/N ratio. Thus no photometry was possible for this filter. Figure\,\ref{ObsImages} shows the HST as well as the best final VLT/NACO images, clearly revealing the increasing separation of the binary components with time.


\subsubsection{Spectroscopy}
Our goal to obtain first spatially-resolved spectra of each component of the binary system was achieved by performing spectroscopy with NACO and its \emph{H}- and \emph{K}-filters (spectral range 1.37 - 1.72\,$\mu$m and 2.02 - 2.53\,$\mu$m) on April 26, 2006. We used  the S27$\_3\_$SH and S27$\_3\_$SK modes, respectively, with a slit width of 0.086\arcsec\, and a resolution of R\,$\sim$1500 and R\,$\sim$1400. A total of 20 spectra (in \emph{H}-band) and 30 spectra (in \emph{K$_{S}$}-band) were obtained in an "abbaabb" nodding pattern along the slit with an exposure time of 60\,s for each frame. The slit was oriented along the Kelu-1\,AB axis, yielding simultaneous spectra of both components at each position. As a telluric calibrator the B4\,IV standard HIP\,067786 was observed immediately after Kelu-1\,AB at a slightly lower airmass. 

In a first step the raw, two-dimensional science spectra were pairwise subtracted to remove the sky and afterwards divided by the median-combined flat-field image. 
Due to the overlapping wings of the barely resolved spectra, we used a custom made IDL routine to perform an iterative fit to extract the individual spectra. This minimum least square fit method turned out to be more robust than standard IRAF reduction procedures leading to an improved S/N ratio. The wavelength calibration was determined based on argon lamp spectra.

To remove the telluric absorption lines and the instrument's relative spectral response function, we used the observed spectra of the B4\,IV telluric standard. After correcting the standard star spectrum for its own hydrogen absorption lines, we divided it by a blackbody spectrum of T$_{\mathrm{eff}}$\,=\,17,050\,K in order to preserve the continuum shape of Kelu-1\,A and B. However, the correction with the standard spectra did not sufficiently correct for all instrumental artifacts. Therefore we calculated an additional response function to correct the remaining slope. The added, individual spectra of Kelu-1\,A and Kelu-1\,B  were first divided by the unresolved near-IR spectra of Kelu-1 from the NIRSPEC Brown Dwarf Spectroscopic Survey (BDSS) \citep{McLean03}, which provided a similar resolution. Then the single Kelu-1 spectra were divided with this derived response function, receiving the finally reduced individual spectra of Kelu-1\,A and B.

In general the achievable S/N ratio is biased by slit losses which are less in \emph{H}- than in \emph{K}-band (wavelength dependent since the FWHM of the AO PSF decreases with increasing wavelength). The \emph{K}-band observation of Kelu-1\,AB was in addition affected by an alignment error of the binary into the slit, which further decreased the S/N ratio. Thus our finally derived S/N ratio in the \emph{H}-band is moderate, though in the \emph{K}-band it is relatively low, especially for the B-component.

\subsection{Astrometry}
\label{Astrometry}
To obtain relative photometry and astrometry of the binary we used our custom-made IDL-based simultaneous PSF-fitting allgorithm \citep{Bouy03}, adapted to our HST/NICMOS and VLT/NACO dataset.  

In order to recover the binary parameters from the first epoch of HST/NICMOS
observations in 1998, we tried to fit the observed light distribution with theoretical
PSFs created with TinyTim, as well as with observed PSFs. A data base of observed NIC1 PSF stars was compiled by querying the HST archive for all observations with NIC1 in F110M obtained within a period of 3 months before or after the Kelu-1 observations. To select the  best-matching PSFs, only observations with NICMOS focus setting "Camera 1", i.e., a PAM (pupil alignment mechanism) focus value close to the value of the Kelu-1 observations were used. This led to a PSF library of 36 individual NIC1 exposures in F110M. Most of the PSFs resulted from the observations of very low mass stars and brown dwarfs in the Pleiades (GO 7952, \citealt{Mart00}), which were well exposed and represent relatively isolated stars.\\
As the photometry obtained from the spatially resolved NACO observations indicates
that both components of Kelu-1 are not variable by more than a few millimagnitudes,
we chose to keep the brightness ratio between the two components fixed
at q\,=\,0.47 for subsequent PSF-fitting, based on the values computed from the NIC1 observations in F108N and F113N obtained in the later epoch.

For the later obtained HST images of the clearly resolved binary, we used a library of 6 different PSFs for a better error estimate: two natural PSFs obtained during previous observations of similar objects of the same program and 4 theoretical PSFs considering different focus settings. The relative photometry and astrometry were measured separately for each of the 4 images per filter. Finally the results were averaged and the uncertainties were calculated from the standard deviation.\\
By contrast, for the VLT data only one reference PSF star was observed in the same night with the same instrumental settings. Therefore all VLT results were based on fitting one single PSF\,. For a better estimate of the accuracy of the results, we repeated the PSF fitting using also the previous PSF star observations. The derived differences are represented in the error calculations.

\subsection{SPITZER Photometry}
The imaging observations with the \emph{Spitzer Space Telescope} (SPITZER) were retrieved from the SPITZER Science Archive. As part of the AOR ID 35 (PI G.Fazio) and AOR ID 3736 (PI B. Goldman) programs, unresolved Kelu-1\,AB imaging data had been obtained on Jan 19, 2004 and Feb 05, 2006, respectively, using the Infrared Array Camera (IRAC). IRAC provides simultaneous broad band images at 3.6, 4.5, 5.8 and 8.0 $\mu$m (also referred to as \emph{channels} 1-4 respectively) with a pixel scale of 1.2\arcsec\,and a FoV of 5.2\arcmin \,x 5.2\arcmin. Using a dichroic beam splitter, two adjacent fields are imaged in pairs (channels 1 and 3; channels 2 and 4). Then the telescope was nodded to image the target in all four channels.

The SPITZER/IRAC data analysis is based on Spitzer Science Center (SSC) pipeline-created mosaic images (with pipeline versions S13.2 and S14.0, respectively), the so-called post-BCD data (pbdc). The photometry was extracted using again the IRAF \emph{phot} routine. As a first step we converted the physical unit of the images from surface brightness into counts by dividing with the conversion factor (found in the header of each image) and multiplying  by the effective integration time of 268s. For the aperture photometry  we chose an aperture size on source of 7 pixels and a sky-radius of 7-14 pixels. This guaranteed to receive all PSF flux and still have a sky area without any object or bad pixel in it. Since the absolute IRAC calibration is based on an aperture with a radius of 10 pixels we applied an aperture correction as listed in the IRAC Data Handbook 3.0. For re-transformation of the net source counts into flux we multiplied with the flux conversion factor and the solid angle for each pixel. Eventually, the instrumental magnitudes were converted into the Vega magnitude scale. The final results are based on the combination of the data sets of both programs and are listed in Table \ref{Mag}.


\section{Results and Discussions}

\subsection{Orbital parameter determination and dynamical mass estimates}
\label{Orbit}
Being clearly resolved as a binary on March 04, 2005 for the first time with a separation of 284.3\,$\pm$\,0.8\,mas and a position angle of 220.9\degr $\pm$\,0.3\degr\,\citep{Gelino}, the separation between the two components increased relatively steady to 366.0\,$\pm$\,0.6\,mas until June 03, 2008. But since the PA only changed by 1.84\degr\,during that time, a reliable new orbit prediction would have been deficient.

Therefore, we also reexamined the HST/NICMOS observations obtained on Aug 14, 1998 (see Sect. \ref{PhotHST} above) and derived a separation of 69.4\,$\pm$\,5.1\,mas and a PA = 22.7\degr $\pm$ 8.2\degr\, as best fitting binary parameters.  As a separation of 65\,mas (or less) corresponds to 70\% of the diffraction limit of HST at 1.1\,$\mu$m, we did not attempt to fit the binary parameters at still longer wavelengths. At an observing wavelength of 1.5\,$\mu$m, 65\,mas are almost exactly half the diffraction limit of HST. Table \ref{Astrom} lists all astrometric results for the Kelu-1\,AB system obtained so far.

\begin{table*}
\caption{Relative astrometric results}
\label{Astrom}
\centering
\begin{tabular}{c c c c c c c }     
\hline\hline
\noalign{\smallskip}
Date & MJD & Telescope/Instrument & Filter & Sep. & PA & References\\
 & & & & [ mas ] & [deg] & \\
\hline
\noalign{\smallskip}
   14\,/\,08\,/\,1998 & 51039.8 & HST/NIC1 & F110M & 69.4 $\pm$ 5.1 & 22.7 $\pm$  8.2 & this work\\
    04\,/\,03\,/\,2005 & 53434.0 & Keck/NIRC2 & K$\arcmin$ & 284.3 $\pm$ 0.8 & 220.90 $\pm$ 0.30 & \citet{Gelino}\\
    01\,/\,05\,/\,2005 & 53491.0 & Keck/NIRC2 & K$\arcmin$& 291.0 $\pm$ 2.0 & 221.20 $\pm$ 0.60 & \citet{Liu}\\
   31\,/\,07\,/\,2005 & 53582.0 & HST/NIC1 & F108N & 299.8 $\pm$ 0.2 & 221.33 $\pm$ 0.04 & this work\\
   28\,/\,04\,/\,2006 & 53853.1&VLT/NACO& K$_\mathrm{S}$ & 320.3 $\pm$ 1.7 & 221.84 $\pm$ 0.1& this work  \\
   26\,/\,05\,/\,2006 & 53881.1&VLT/NACO & L$\arcmin$ & 322.7 $\pm$ 1.9 & 221.86 $\pm$ 0.1 & this work\\
   12\,/\,08\,/\,2006 &53959.9 &VLT/NACO & K$_\mathrm{S}$ & 330.6 $\pm$ 5.0 & 222.24 $\pm$ 0.1 & this work\\
   22\,/\,02\,/\,2007 &54153.4 &VLT/NACO & K$_\mathrm{S}$ & 341.5 $\pm$ 1.5 & 222.39 $\pm$ 0.1 & this work\\
   15\,/\,05\,/\,2007 &54235.9 &VLT/NACO & K$_\mathrm{S}$ & 345.3 $\pm$ 1.6 & 222.48 $\pm$ 0.1 & this work\\
   13\,/\,03\,/\,2008 &54538.8 &VLT/NACO & K$_\mathrm{S}$ & 362.8 $\pm$ 0.6 & 222.63 $\pm$ 0.1 & this work\\
   09\,/\,05\,/\,2008 &54595.7 &VLT/NACO & K$_\mathrm{S}$ & 364.5 $\pm$ 0.5 & 222.67 $\pm$ 0.1 & this work\\
   03\,/\,06\,/\,2008 &54621.5 &VLT/NACO & K$_\mathrm{S}$ & 366.0 $\pm$ 0.6 & 222.74 $\pm$ 0.1 & this work\\
\noalign{\smallskip}
\hline
\end{tabular}
\end{table*}

Compared to the very first orbital predictions by \citet{Liu}  and \citet{Gelino}, our nine additional data points allow us to improve the determination of the orbital parameters substantially. Based on all astrometric measurements obtained so far, the best orbital solution was determined with our own IDL-based algorithm (see \citet{Koeh} for a detailed description of the method). The procedure consists of two steps:  First, a grid search in period $P$, eccentricity $e$ and time of periastron $T_0$ is carried out, while Singular Value
Decomposition is used to solve the linear equation system for the four Thiele-Innes constants.  From the Thiele-Innes constants, the remaining orbital elements (semi-major axis $a$, argument of periastron $\omega$, position angle of the line of nodes $\Omega$, and inclination $i$) are computed.  The result of the grid search is $\chi^2$ as function of $P$ and $e$, which helps to assess the quality of the fit, and the confidence limits of the fit parameters.

In the second step, all 7 orbital elements are fit simultaneously by a Levenberg-Marquardt algorithm.  This further improves the fit by interpolating between grid points and ensures that all 7 elements are treated equally in the final fit.  To avoid that the algorithm converges on a local instead of the global minimum, we decided to use {\em all\/} orbits resulting from the grid-search as starting points and carried out $148\times100$ runs of the Levenberg-Marquardt algorithm (the grid spans 148 periods and 100 eccentricities).

To derive estimates of the confidence intervals for the parameters, we studied the $\chi^2$ function around its minimum.  To find the confidence interval for one parameter (for example $T_0$), we perturbed this parameter away from the minimum in $\chi^2$ and optimized all the other parameters.  Any perturbation of a parameter will of course lead to a larger $\chi^2$.  The range in $T_0$ within which $\chi^2(T_0) - \chi^2_{\rm min} < 1$ defines the 68\,\% confidence interval for $T_0$.  This interval is usually not symmetric around the $T_0$ of the best fit, therefore we list in Table~\ref{OrbPar} separate limits for positive and negative perturbations.

Finally, the total system mass is computed from the semi-major axis and orbital period through Kepler's third law. To convert the semi-major axis from arcsec into AU as well as to compute the mass in M$_\mathrm{Jup}$\, we used a distance of 18.7 $\pm$ 0.7\,pc as determined by \citet{Dahn}.

\begin{figure*}[t!]
  \centering
       \includegraphics[width=12cm]{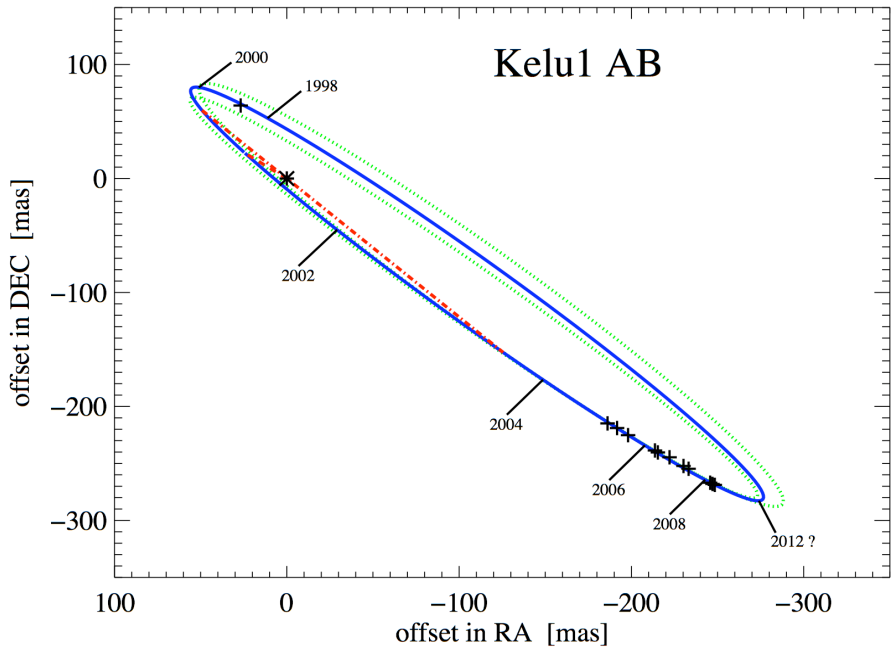}
  \caption{Orbit of the Kelu-1\,AB system. The star indicates the position of Kelu-1\,A  while the measurements  for the relative motion of Kelu-1\,B around Kelu-1\,A are indicated by crosses. Measurement errors are comparable to or smaller than the plotted symbols. The resulting best orbit fit is represented by the solid, blue curve and the green, dotted lines show the orbits at the 90\% confidence limit (with periods of 32 and 46 years). The red, dash-dotted line marks the line of nodes and the dashed line indicates the position of the periastron. In addition the positions on January 1$^{st}$ for the years 1998-2008, as well as the predicted position for 2012 are marked. }
   \label{OrbitPred}
  \end{figure*}

We applied this IDL code to the relative astrometric data as given in Table \ref{Astrom} (hereafter data set N$^\mathrm{o}$1) and found a global minimum reduced $\chi^2$ of 1.3. The best-fit results are summarized in Table \ref{OrbPar}, including the uncertainties of each fitted parameter, and the orbital solution is illustrated in Fig. \ref{OrbitPred}. The orbit is quite eccentric ($e$ = 0.82\,$\pm$\,0.10 ) with an inclination of 84.9\degr  \,$^{+ 1.0}_{ -2.0}$\,, i.e.\ seen almost edge-on. The period of $38^{+8}_{-6}$\,years and the semi-major axis of $6.4^{+2.4}_{-1.3}$\,AU yield a total system mass of
$177^{+113}_{-55}\,M_{\rm Jup}$. This is the first purely dynamical mass estimate for the Kelu-1\,AB system with no evolutionary models involved in the calculations. 

%
\begin{table}[b]
\caption{Parameters for the best orbital solution of Kelu-1\,AB}
\label{OrbPar}
\centering
\begin{tabular}{l c }     
\hline\hline
\noalign{\smallskip}
Parameter &  Value\\
\hline
\noalign{\smallskip}
System mass $M_S$ [$M_\mathrm{Jup}$]                          & $  177$  $^{+113}_{-55}$\\ [1.5ex]
Period $P$ [years]                                  & $     38$  $^{+   8}_{   -6}$\\[1.0ex]
Semi-major axis $a$ [mas]                         & $    339$  $^{+ 129}_{- 66}$\\ [0.5ex]
\hspace{16.5ex}  [AU]          & $    6.4$ $^{+ 2.4}_{ -1.3}$\\ [1.0ex]
Eccentricity $e$                                  & $   0.82$  $^{+0.10}_{-0.10}$\\ [1.0ex]
Argument of periastron $\omega$ [$^\circ$]          & $   57.8$  $^{+15.0}_{-20.0}$\\[1.0ex]
P.A. of ascending node $\Omega$ [$^\circ$]         & $   39.4$  $^{+ 2.0}_{
-2.0}$\strut\\[1.0ex]
Inclination $i$ [$^\circ$]                         & $   84.9$  $^{+ 1.0}_{ -2.0}$\\[1.0ex]
Date of periastron $T_0$                          & $2452079$  $^{+ 200}_{ -400}$\\[0.5ex]
                                                  & (Jun 18, 2001)\\ [1.5ex]
reduced $\chi^2$                                  & $    1.3$\\
\noalign{\smallskip}
\hline
\end{tabular}
\end{table}


%

This mass prediction reveals a slightly higher total system mass than one would expect for a BD--BD binary system, where an upper mass limit of 72 -- 75 $M_\mathrm{Jup}$ for brown dwarfs (the substellar mass-boundary) would result in a maximum total system mass of $\sim$\,150 $M_\mathrm{Jup}$. Though, within the uncertainties of our estimate a pure BD--BD binary is still possible.  The known presence of Li absorption in the unresolved optical spectrum \citep{Ruiz97, Kirk99} indicates that at least one component must have a mass $\la$ 65 $M_\mathrm{Jup}$  which is the lithium-burning limit \citep{Rebolo, Basri98_2}. If the system consists of two objects, this would imply that the primary has a mass of $\approx$\,110 $M_\mathrm{Jup}$, which in turn would correspond to a main-sequence spectral type of M8V. This is in contradiction to its early L spectral type as it has been determined both from integrated spectra \citep{Kirk99, Dahn} as well as in $\S$\,\ref{SpecRes} below from the spatially resolved spectroscopic observations of Kelu-1\,A and B.

\subsubsection{Comparison to previous orbit predictions}
Since the HST/NICMOS observation from 1998 is obviously a key component to fit the orbit, we compared our binary parameters for this observation with those obtained by \citet{Gelino}. They derived a separation of 45\,$\pm$\,18\,mas and a PA = 38.0\degr \,or 218.0\degr \,$\pm$ 11.9\degr\, (due to their 180\degr \,ambiguity in position angle due to the unresolved objects). Within the uncertainties, these values are in good agreement with our measurements. In an additional check we analyzed what orbit solutions we would get with their slightly different result. We applied our IDL code on two additional data sets which both contained the same values as data set N$^\mathrm{o}$1 except for the very first measurement. In data set N$^\mathrm{o}$2 we replaced the values of the August 1998 observation by a separation of 45\,$\pm$\,18\,mas and a PA of 38.0\degr \, while data set N$^\mathrm{o}$3 used a PA of 218.0\degr \,. The best-fitting orbital solution for data set N$^\mathrm{o}$3 results in a physically unacceptable total system mass of 108 \emph{solar masses}. If we consider only orbital models with a total system mass $\la$ 230 $M_\mathrm{Jup}$ (assuming a possible triple system), we find a minimum reduced $\chi^2$ of 1.5, which is higher than that for data set
N$^\mathrm{o}$1. Therefore we exclude the PA of 218\degr\, as a possible first data point.  The result from the best-fit solution for set N$^\mathrm{o}$2 yields, without any constraints, a system mass of 97$^{+105}_{-23}$ M$_\mathrm{Jup}$, period of $28^{+6}_{-10}$ years and semi-major axis of $4.3^{+0.5}_{-0.2}$ AU. These values are all slightly smaller, though within the uncertainties still consistent with the result for data set N$^\mathrm{o}$1. However, since its minimum reduced $\chi^2$ of 1.4 is still higher
than the one for set N$^\mathrm{o}$1 we stay with our own astrometric result for the first HST/NICMOS data point.  In addition, this ensures a consistent data reduction method throughout the whole monitoring program.

Comparing the final results in Table \ref{Astrom} with the previously published orbit parameter predictions by \citeauthor{Liu} and \citeauthor{Gelino} we first exclude their assumed circular orbit, since with an eccentricity $\le$\, 0.01 this would result in unphysical parameters like a dynamical total system mass of 1.06 M$_\odot$ or a reduced $\chi^2 > 3$. Nevertheless, our derived inclination of 84.9$^{+ 1.0}_{ -2.0}$\degr \,agrees well with the limiting values of 81 $\le$ \emph{i} $\le$ 90\degr\, reported by \citeauthor{Gelino}. With a total system mass of 177\, $M_\mathrm{Jup}$, our prediction is higher than the mass estimates of $\sim$\,120 $M_\mathrm{Jup}$ and \,$\sim$\,115 $M_\mathrm{Jup}$ by \citeauthor{Liu} and \citeauthor{Gelino} respectively. However, their calculations were purely based on photometric results compared with theoretical models and a still uncertain age estimation for the system.

Considering a possible third component in the system (as derived from the results in $\S$\,\ref{KeluA_spec} below), would allow all objects to have component masses $\le$ 95 $M_\mathrm{Jup}$, and hence spectral types of L0 or later. This would resolve the above mentioned discrepancy between a system mass of $\sim$180 $M_\mathrm{Jup}$, spectral types of L0 and later, and the fact that at least one of the components must have a mass less than 65 $M_\mathrm{Jup}$ as indicated by the presence of Lithium in the combined spectra.

All together we showed a possible solution in reasonable agreement with the observations, but most likely not the final one,  since the fraction of the orbit observed so far is too small for a reliable and consistent orbit determination. For a robust orbital solution, more observations are required, at least up to the turning point in the (projected) orbit expected around the year 2012 (see Fig. \ref{OrbitPred}). Therefore further investigations, including the measurement of the radial velocities of the system components, will be done with our ongoing long-term monitoring program. 
  \begin{figure*}[t!]
  \centering
    \setlength{\unitlength}{1cm}
    \begin{minipage}[t]{8.9cm}
       \includegraphics[width=\textwidth]{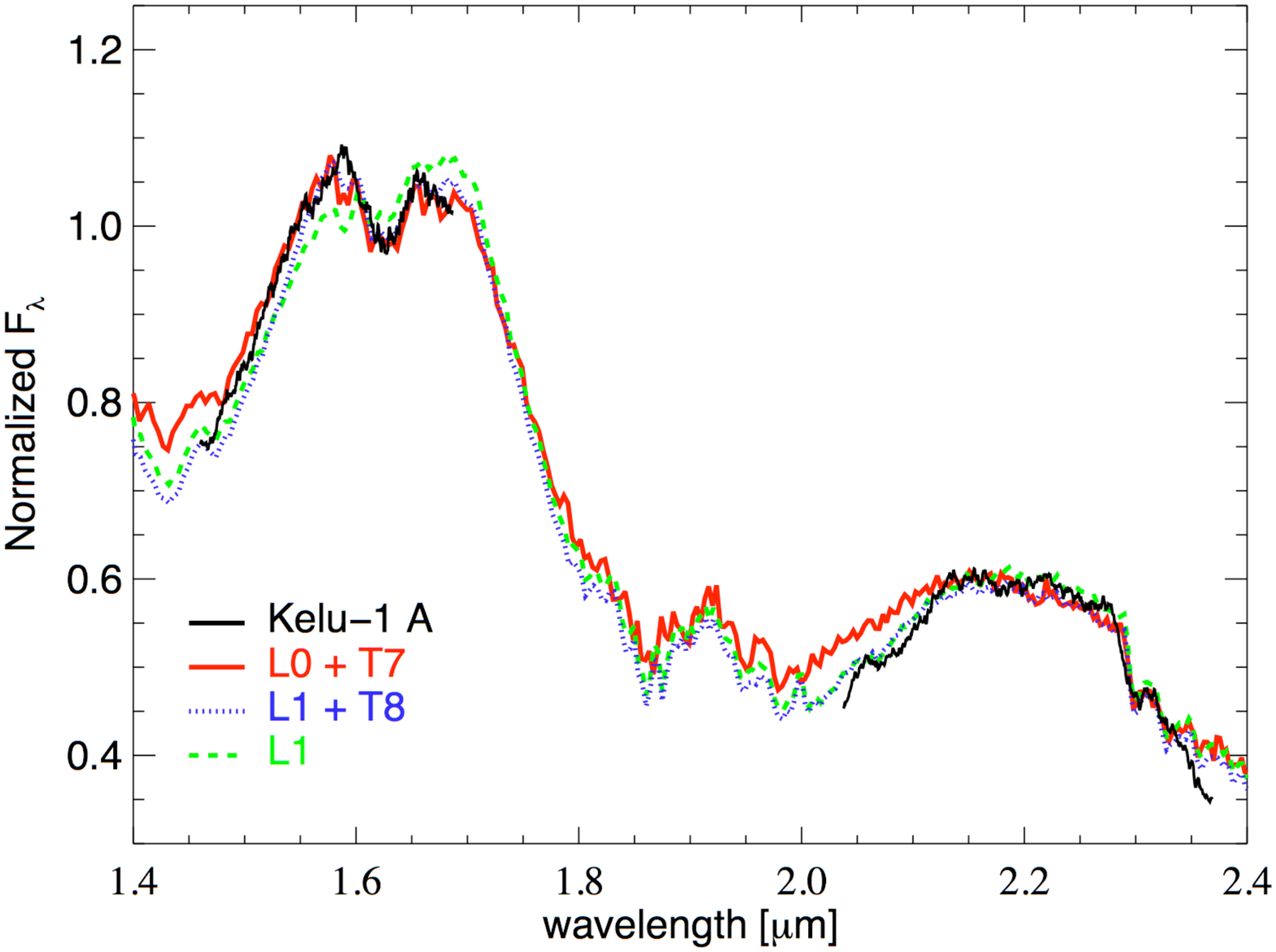}
     \end{minipage}
    \begin{minipage}[t]{8.9cm} 
       \includegraphics[width=\textwidth]{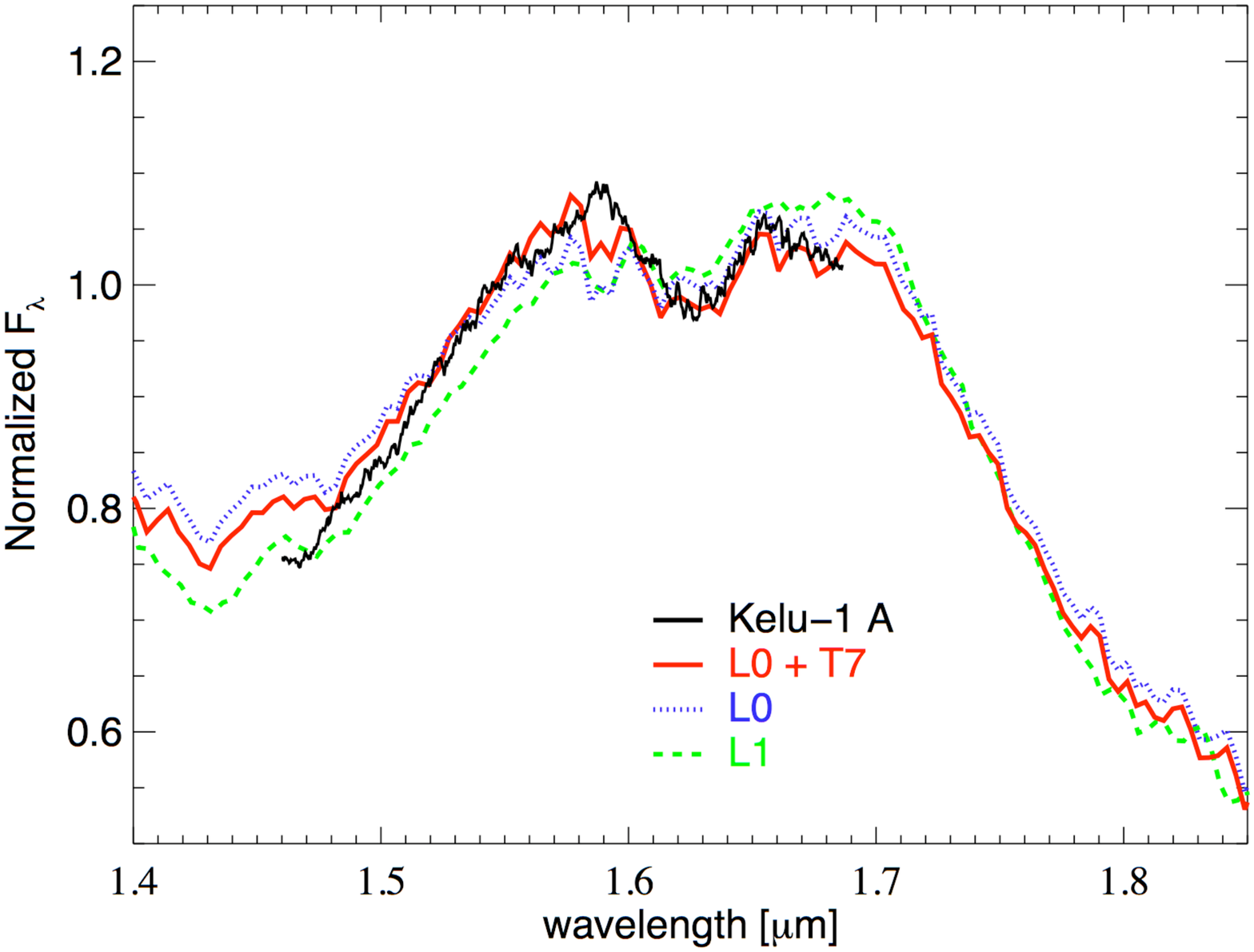}
      \end{minipage}
  \caption{ \emph{Left}: \emph{H}- and \emph{K}-band spectrum of Kelu-1\,A (smoothed to template resolution, solid line) compared to the best-fit binary composites of 2M2107-0307 (L0) combined with 2M0727-1710 (T7) (red, solid) and  2M1439-1929 (L1) with 2M0415-0935 (T8) (blue, dotted). In addition the single object spectrum of 2M2107-0307 (L0) (green, dashed) is plotted for comparison. 
\emph{Right}: close-up of the \emph{H}-band and the peculiar feature starting at 1.6\,$\mu$m. While the single L0 (blue, dotted) and L1 (green, dashed) show a plateau shape from 1.55 -- 1.64\,$\mu$m due to the FeH absorption band heads, the L0$\,+\,$T7 binary composite (red, solid) reproduces the deep depression to a much higher accuracy.}
   \label{SpecKeluA}
  \end{figure*}
%

\subsection{Spectral types}
\label{SpecRes}
\begin{figure*}[t!]
  \centering
    \setlength{\unitlength}{1cm}
    \begin{minipage}[t]{8.9cm}
       \includegraphics[width=\textwidth]{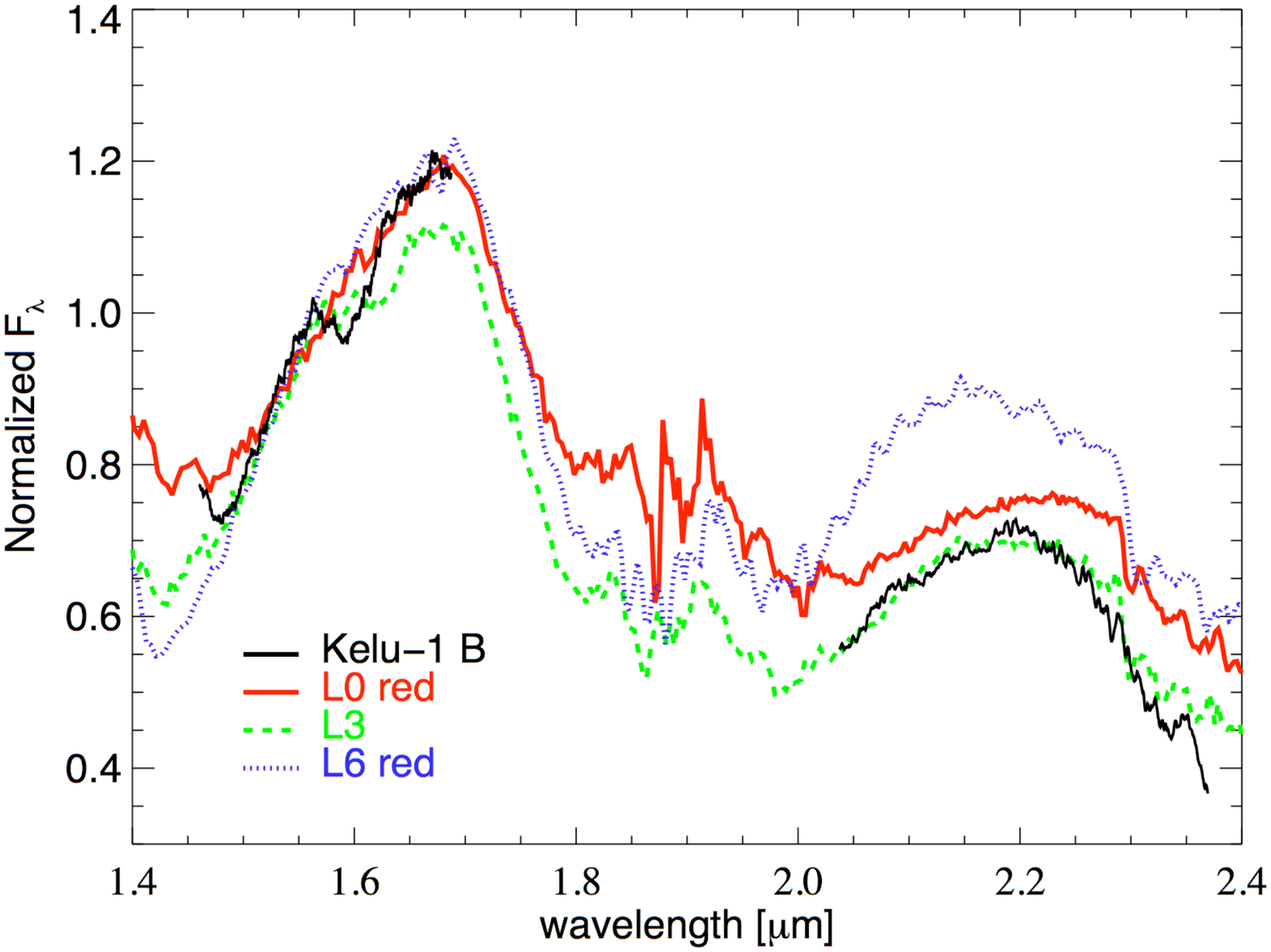}
     \end{minipage}
    \begin{minipage}[t]{8.9cm} 
       \includegraphics[width=\textwidth]{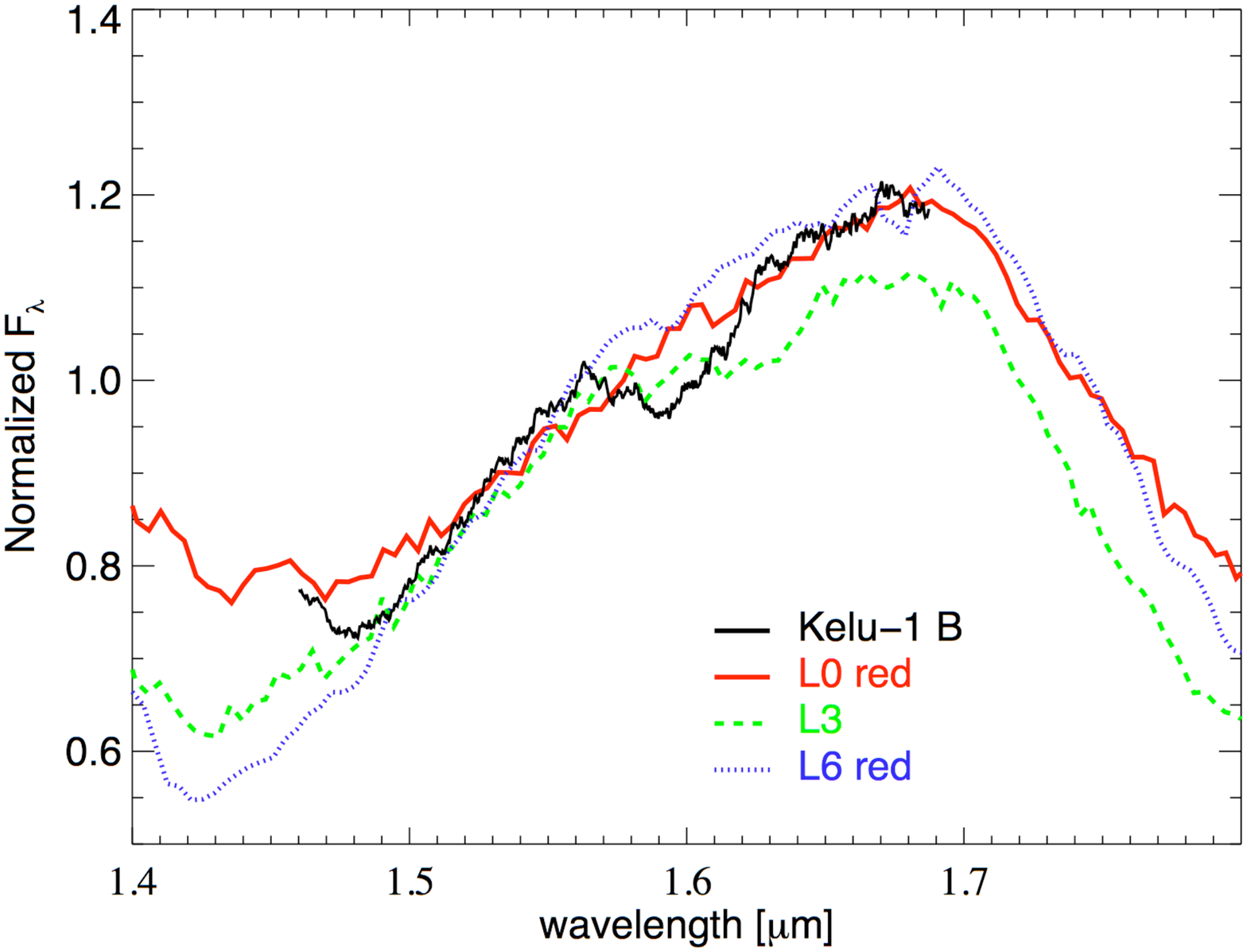}
      \end{minipage}
  \caption{\emph{Left}: \emph{H}- and \emph{K}-band spectrum of Kelu-1\,B (smoothed to template resolution, solid line) compared to the best-fit spectra of the red L dwarfs 2M0141-4633 (L0\,pec, red solid line) and 2M0103$+$1935 (L6, blue dotted line), as well as the field L dwarf 2M1506-1321 (L3, green dashed line).
\emph{Right}: close-up of the \emph{H}-band with the same objects as on the left. The distinct FeH absorption dip around 1.58\,$\mu$m is present in the L3 spectrum, but not in the L0\,pec or L6 spectra, while the overall slope better fits the red objects.}
   \label{SpecKeluB}
  \end{figure*}

The extracted and reduced \emph{HK}- spectra of Kelu-1\,A and Kelu-1\,B are shown in Figs. \ref{SpecKeluA} and \ref{SpecKeluB}, respectively. To determine the spectral type of each component we compared them with low-resolution spectra of 5 late M dwarfs (M8-M9.5) and 39 field L dwarfs (L0-L9) from the SpeX prism data archive\footnote{http://www.browndwarfs.org/spexprism}. All spectra were normalized to the \emph{H}-band. In addition the Kelu-1 spectra were smoothed to the template resolution and scaled by a factor of 0.95 to eliminate the normalization bias.

\subsubsection{Kelu-1\,A}
\label{KeluA_spec}
The general \emph{H}-band slope of Kelu-1\,A best fits between the optically classified L0 2MASSI J2107316-030733 (hereafter 2M2107-0307, \citealt{Burgasser04}) and the L1 optical standard 2MASSW J1439283+192915 (hereafter 2M1439$+$1929, \citealt{Burgasser04}). In the \emph{K}-band the overall shape is clearly very similar to that of 2M1439$+$1929 with the same strong CO absorption band heads at 2.3\,--\,2.4\,$\mu$m and, at the short-wavelength end of the \emph{K}-band, a steeper slope than in the L0 spectrum, due to stronger H$_{2}$O absorption. This suggests a spectral type of L0.5 $\pm$ 0.5 which is slightly earlier than the derived spectral type of L1.5 -- L3 from \citet{Liu} whose estimates were only based on \emph{JHK} colors and absolute magnitudes. 
However, the \emph{H}-band spectrum of Kelu-1\,A shows a pronounced feature, a deep, broad depression from 1.6 -- 1.65\,$\mu$m. This feature is slightly offset from the 1.55 -- 1.64\,$\mu$m FeH absorption band heads which produce a flat plateau in this wavelength range and are usually present in early to mid type L dwarfs \citep{Cushing03}. Instead, the 1.6\,$\mu$m depression corresponds much more to the CH$_{4}$ absorption band which starts to appear at the L/T transition and strengthens through the T sequence and is therefore very unusual for an early L dwarf. In addition there is no indication of the 2.2\,$\mu$m CH$_{4}$ absorption line, another defining feature for the T dwarfs.

Recently the same feature has been observed in 3 formerly classified single, early to mid-type L dwarfs (2MASS J05185995-2828372, \citealt{Cruz04};  SDSS J080531.84+481233.0, \citealt{Burgasser07}; 2MASS J03202839-0446358, \citealt{Burgasser08}). By combining an early to mid L dwarf spectrum with an early to mid T dwarf spectrum, these objects were reinterpreted to be unresolved spectroscopic binaries, with the depression being very well reproduced by a combination of the FeH absorption in the L dwarf primary and the CH$_{4}$ absorption of the T dwarf secondary. Two of the targets have been resolved by observations since then: 2MASS 0518-2828 with high-resolution imaging \citep{Burgasser06_1} and 2MASS 0320-0446 with high-resolution spectroscopy to be a single-lined spectroscopic binary (SB1) \citep{Blake08}.

To probe if we can also reproduce the depression in the Kelu-1\,A spectrum including the overall shape of the \emph{H}- and \emph{K}-band, we created a large set of binary spectral templates by combining late M-dwarf\,/\,early L dwarf spectra (M8--L4) with early to late T dwarf spectra (T2--T8) from the SpeX prism data archive. We selected the SpeX data archive since it is the only archive which provides a continuous and large set of spectral types from late M-dwarfs to the latest T dwarfs in the \emph{H}- and \emph{K}-band and, therefore, provides the most homogenous set for assembling all the different binary templates. After normalization we calculated the RMS between the science spectrum and the binary templates to determine the best matches. The results with the least deviation are displayed in Fig. \ref{SpecKeluA} (left). The combination of  2M2107-0307 (L0) and  2MASS J07271824+1710012 (T7\,standard) (2M0727$+$1710,  \citealt{Burgasser06_2}) best fits the slope and depression in the \emph{H}-band. 
%
Whereas the smallest RMS in the \emph{K}-band is achieved with the combined spectrum of 2M1439-1929 (L1) and 2MASS J04151954-0935066 (T8 standard) (2M0415-0935, \citealt{Burgasser04}) with a better match of the steeper slope at the short-wavelength end of the \emph{K}-band spectrum due to stronger H$_{2}$O absorption.  On the right side of Fig. \ref{SpecKeluA} we show a close-up of the \emph{H}-band and the peculiar feature starting at 1.59\,$\mu$m. One can clearly see that neither the single best-fit L0 nor the L1 match the deep depression while the L0$+$T7 binary template reproduces the whole spectra to a much higher accuracy.
Beside those best-fit comparisons there were several other matches with similar good results among the binary templates with secondaries of type T6.5 and later. Including the small difference between the best match with an L0 primary in \emph{H}-band and an L1 primary in \emph{K}-band we derive a component spectral type of L0.5 $\pm$ 0.5 for Kelu-1\,Aa and T7.5 $\pm$ 1 for Kelu-1\,Ab.


\subsubsection{Kelu-1\,B}
\label{KeluB_spec}
A precise spectroscopic classification of Kelu-1\,B is even more challenging. The object shows an \emph{H}-band continuum notably different from the usual plateau-shape of "normal" early to mid-type L dwarfs (see Fig. \ref{SpecKeluB}). Its sharply peaked, triangular-shaped continuum resembles much more that of the young, red L0pec 2MASS J01415823-4633574 (2M0141-4633, \citealt{Kirk06}) and the red L6 2MASSI J0103320+193536 (2M0103$+$1935, \citealt{Cruz04}). While the distinct dip around 1.58\,$\mu$m in the Kelu-1\,B spectrum (caused by two prominent FeH absorption bands) is not visible in the red spectrum of the L0pec and L6 dwarf, it is in turn present only slightly weaker in  "normal"  L dwarf spectra.

By contrast the \emph{K}-band spectrum of Kelu-1\,B does not fit any of the two red objects. While its overall flux distribution fits relatively well that of the L3 optical standard  2MASS J15065441+1321060 (2M1506$+$1321, \citealt{Burgasser07_2}), it again reveals a  triangular-shaped continuum with a slightly redward shifted \emph{K}-band flux. In addition the CO absorption bands are somewhat weaker as in the L3 dwarf.

For comparison in the \emph{H}-band, only the above mentioned two unusual red L dwarfs were available in the archive. Therefore, we used the surface gravity independent H$_{2}$O indices in the short-wavelength end of the \emph{H}-band (1.49--1.56\,$\mu$m) from \citet{Allers07} for a better estimation of the spectral type. In a first attempt we applied this  H$_{2}$O index to the L0pec 2M0104-4633 and obtained a spectral type of M9.7 $\pm$ 0.5,  which is fully consistent with the L0 classification of \citet{Kirk06}. Using the same indices for Kelu-1\,B we receive a spectral type of L2.6 $\pm$ 1.5 where the uncertainty includes the scatter of the index relation and the error of our index measurements. Using the results from our fits and calculations we thus adopt an overall spectral type of L3\,pec $\pm$1.5 for Kelu-1\,B.

Similar peaked \emph{H}-band spectra and unusual red near-IR colors have been so far mainly detected as characteristics for young ultracool dwarfs (\citealt{Luhm04}; \citealt{Kirk06}; \citealt{Allers07}; \citealt{McEl07}; \citealt{Metchev08}). The effect can be explained by two possible scenarios which are both caused by lower gravity and lower atmospheric pressure: (i) a diminishing collision-induced absorption (CIA) of H$_{2}$ in the \emph{H}- and \emph{K}-band \citep{Kirk06}, or (ii) increased water vapor absorption on each side of the \emph{H}-band peak (\citealt{Luhm04}; \citealt{Allers07}).  
Another possible explanation for the peculiar shape is a higher metallicity which implies a higher opacity caused by greater dust production. This again results in an overall redder spectrum with weaker alkali (Na, \ion{K}{i}\,) and molecular (FeH and H$_{2}$O) absorption lines and a red-shifted \emph{K}-band peak.

\begin{table*}[t!]
\begin{minipage}[t]{\columnwidth}
\caption{Apparent and absolute magnitudes of Kelu-1\,AB}
\label{Magnitudes}
\centering
\begin{tabular}{l c c c c c c c }     
\hline\hline
\noalign{\smallskip}
 &&Kelu-1\,AB&& \multicolumn{2}{c}{Kelu-1 A} & \multicolumn{2}{c}{Kelu-1 B}\\
\raisebox{1.5ex}[1.5ex]{Telescope/Instrument} & \raisebox{1.5ex}[1.5ex]{Filter} & m [mag] & \raisebox{1.5ex}[1.5ex]{$\Delta$ mag} &m [mag] & M [mag]& m [mag] &  M [mag] \\
\noalign{\smallskip}
\hline
\noalign{\smallskip}
  HST/NIC1 & F108N  & &0.83 $\pm$ 0.02 & 14.56 $\pm$ 0.02& 13.20 $\pm$ 0.08 &15.31$\pm$ 0.02 & 13.95 $\pm$ 0.08\\
  & F113N & &0.82 $\pm$ 0.02 &14.28 $\pm$ 0.02 & 12.92 $\pm$ 0.08 & 15.01$\pm$ 0.03 & 13.65 $\pm$ 0.09\\
  \noalign{\medskip}
  2MASS $^{\mathit{a}}$& J  &13.414 $\pm$ 0.026 &&&&\\
  VLT/NACO&  J  && 0.82 $\pm$ 0.09& 13.83 $\pm$ 0.08 & 12.47 $\pm$0.11 &14.65 $\pm$ 0.12 & 13.29 $\pm$ 0.14 \\
   \noalign{\medskip}
   2MASS $^{\mathit{a}}$  & H  &12.392 $\pm$ 0.025&&&& \\
   VLT/NACO &  H   &&  0.51 $\pm$ 0.02 & 12.92$\pm$ 0.03 & 11.56 $\pm$ 0.09 &13.44 $\pm$ 0.04 & 12.08 $\pm$ 0.09\\
    \noalign{\medskip}
    2MASS $^{\mathit{a}}$& K$_\mathrm{S}$  &11.747 $\pm$ 0.023&&&&\\ 
   VLT/NACO& K$_\mathrm{S}$  & & 0.39 $\pm$ 0.01& 12.32 $\pm$ 0.02 & 10.96 $\pm$ 0.08 & 12.72 $\pm$ 0.03 & 11.36 $\pm$ 0.09\\
   \noalign{\medskip}
   UKIRT/IRCAM $^{\mathit{b}}$ & L$\arcmin$  & 10.78 $\pm$ 0.15 &&&&\\
   VLT/NACO &  L$\arcmin$  & & 0.24 $\pm$ 0.01& 11.42 $\pm$ 0.15 & 10.06 $\pm$ 0.17 & 11.66 $\pm$ 0.15 & 10.30 $\pm$ 0.17\\
   \noalign{\medskip}
    Spitzer/IRAC& 3.6 $\mu$m & 10.90 $\pm$ 0.03 &&&&\\
    & 4.5 $\mu$m  & 10.85 $\pm$ 0.04 &&&&\\
    & 5.8 $\mu$m  & 10.68 $\pm$ 0.07&&&&\\
    & 8.0 $\mu$m  & 10.55 $\pm$ 0.05 &&&&\\
    
\noalign{\smallskip}
\hline
\end{tabular}
\begin{list}{}{}
\item[$^{\mathit{a}}$] from \citet{Cutri}
\item[$^{\mathit{b}}$] from \citet{Leggett02}
\end{list}
\end{minipage}
\end{table*}

While Kelu-1\,B also shows slightly redder J-K$_\mathrm{S}$ colors (see $\S$\,\ref{Mag} below), the triangular-shaped \emph{H}- and \emph{K}-band spectra can not be explained by low gravity and thus implying a young age, since Kelu-1\,A, the other member of the coeval binary system, shows (even with a possible companion) the "normal" behavior of an early L dwarf and no spectral evidence of a young ultracool object. 
Recently \citet{Looper08} identified two new peculiar L dwarfs which showed the same triangular-shaped \emph{H}- and \emph{K}-band spectra as Kelu-1\,B. For one of those objects, the L6.5\,pec 2MASS J21481628+4003593 they suggest that the effect may arise from a metal rich atmosphere rather than from low gravity, since its large tangential velocity does not indicate an unusual young age.

Since low gravity as well as higher metallicity result in weaker alkali and FeH absorption, especially in the red-optical and J-band spectra, further investigations on resolved Kelu-1\,B spectra in these wavelength ranges will be necessary for a better spectral characterization.


\subsection{Magnitudes and Colors }
\label{Mag}
We calculated the individual magnitudes of the components in the J, H, K$_\mathrm{S}$ and L$\arcmin$ filters from the measured binary flux ratios (see \S\,\ref{Astrometry}) and the integrated 2MASS \citep{Cutri} and UKIRT \citep{Leggett02} magnitudes of the unresolved Kelu-1. 
Even if the 2MASS and VLT/NACO near-IR filters are not exactly identical we did not apply any correction factor, given that the spectral energy distributions (SEDs) of the Kelu-1\,AB components are so similar that the flux ratio should not be noticeably affected by the difference in these two photometric systems. For the UKIRT L$\arcmin$ filter one has to consider that it cuts on slightly bluer wavelength than the VLT/NACO L$\arcmin$ filter. 
However, \citet{Leggett02} showed that a similar difference between the UKIRT and MKO filter in this particular wavelength regime is only significant for late L dwarfs and T dwarfs when the CH$_{4}$ absorption at 3.3\,$\mu$m starts to play a role. For early L dwarfs they claim the effect to be less than the uncertainties in the published magnitudes. Thus we used no correction factors between the UKIRT L$\arcmin$ and NACO L$\arcmin$ filter system. 

Table \ref{Magnitudes} summarizes the photometric results. For the determination of the absolute magnitudes we used a distance of 18.7\,$\pm$\,0.7 pc  from \citet{Dahn}. Our error estimations include the uncertainties in the flux ratios, the corresponding uncertainties of the "unresolved" magnitudes, as well as the error of the distance. For comparisons with J, H and K magnitudes and spectral type relations derived in the MKO photometric system, we used the transformation equation for 2MASS into MKO of \citet{Stephens04}. 
  \begin{figure}[h!]
  \centering
       \includegraphics[width=6cm]{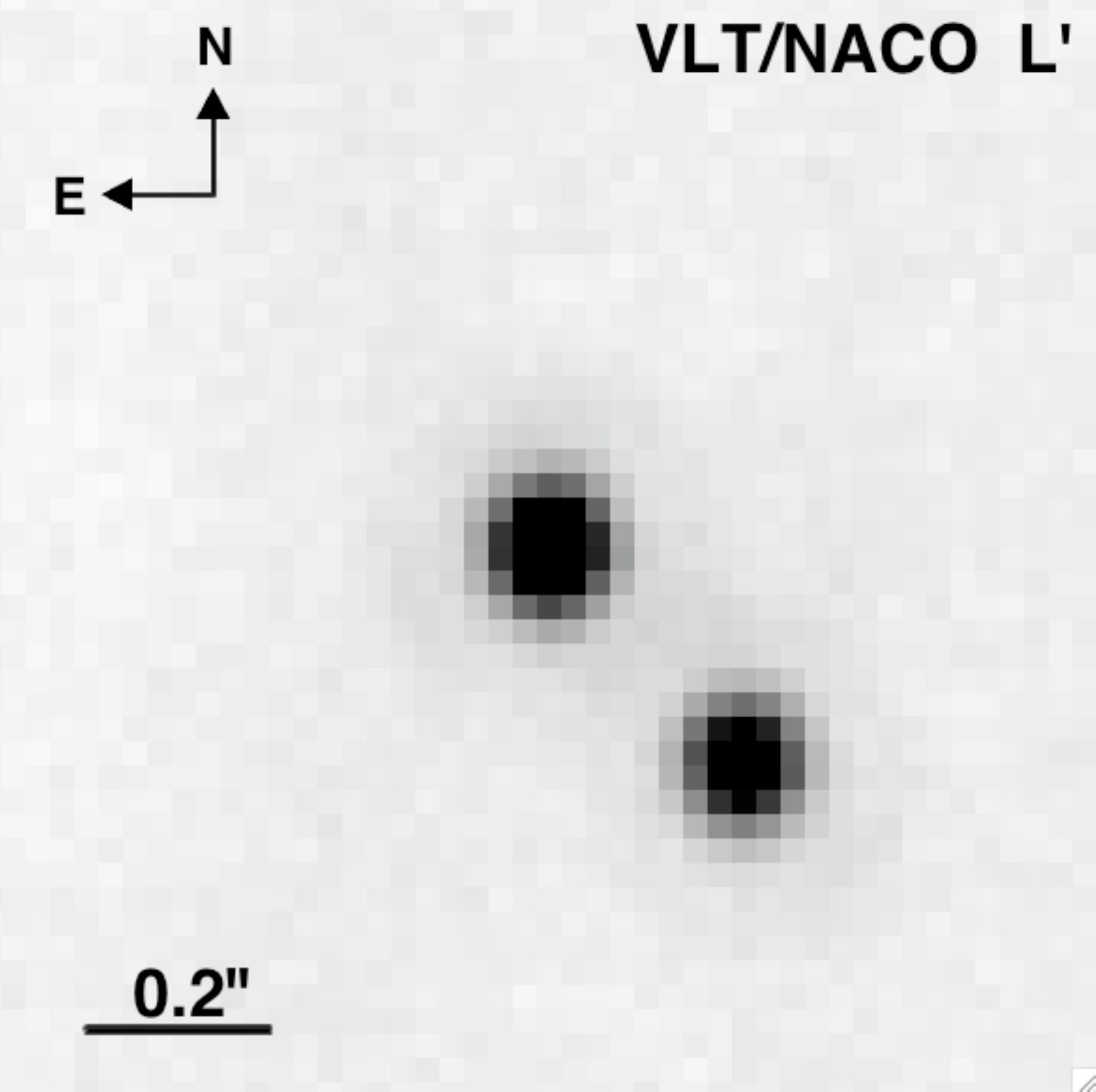}
  \caption{Kelu-1\,AB image achieved with VLT/NACO on 26. May 2006. This is the first resolved L$\arcmin$-band image of a brown dwarf binary system.}
   \label{L_band}
  \end{figure}

%

The photometric monitoring in the K$_\mathrm{S}$-band over the past 2 years revealed no evidence of photometric variability within the errors for Kelu-1\,A or Kelu-1\,B. Compared to the values of \citet{Liu}, our individual magnitudes in H and K$_\mathrm{S}$-band are in good agreement within the errors, whereas our J-band magnitude for Kelu-1\,B is slightly fainter. This is likely due to in general lower AO performance in J-band, resulting in a considerably worse Strehl ratio and leading to broader PSF wings, which were hence not separated enough for an accurate flux ratio calculation. We took this into account by assuming a larger error. While the resulting J-K$_\mathrm{S}$ and J-H colors for both components are at the red end, both H-K$_\mathrm{S}$ colors fit very well into the color range for spectral types of L0-L1 and L3-L4 (see \citealt{Cushing08}) respectively.  The red colors of Kelu-1\,B  correspond to the redder \emph{H}- and \emph{K}-band spectra reported and discussed in \S\,\ref{KeluB_spec}. 

We note that our derived  L$\arcmin$-band photometry with L$\arcmin_\mathrm{A}$\,=\,11.42\,$\pm$\,0.15\,mag and L$\arcmin_\mathrm{B}$\,=\,11.66\,$\pm$\,0.15\,mag are the first resolved L-band magnitudes for any L dwarf binary so far (see Fig. \ref{L_band}). While the unresolved L$\arcmin$ photometry was as overluminous as in the other near-IR bandpasses for a spectral type of L3 \citep{Goli04_1}, our resolved (K$_\mathrm{MKO}$\,-\,L$\arcmin$)$_\mathrm{A}$ = 0.88\,$\pm$\,0.15 and (K$_\mathrm{MKO}$\,-\,L$\arcmin$)$_\mathrm{B}$ = 1.03\,$\pm$\,0.15 colors, as well as the absolute L$\arcmin$ magnitudes are consistent with a spectral type of L0.5 and L3, respectively (as in \S\,\ref{SpecRes}), derived from the color-spectral type and magnitude-spectral type relations in \citet{Goli04_1}. 

To get an idea of the photometric values of the potential third component in the Kelu-1\,AB system, we computed the J, H and K$_\mathrm{S}$ magnitudes from the Kelu-1\,A magnitudes. We considered the allowed flux ratios of Aa and Ab in a filter, as well as the latest 2MASS color-spectral type relations in \citet{Cushing08} under the assumption of spectral types $\sim$\,L0.5 and $\sim$\,T7 as derived in \S\,\ref{SpecRes}. 
The primary Kelu-1\,Aa would be only slightly fainter than the composite source with J$_\mathrm{Aa}$\,=\,13.89, H$_\mathrm{Aa}$\,=\,12.95 and (K$_\mathrm{S}$)$_\mathrm{Aa}$\,=\,12.33 while Kelu-1\,Ab is notably fainter: J$_\mathrm{Ab}$\,=\,16.99, H$_\mathrm{Ab}$\,=\,16.95 and (K$_\mathrm{S}$)$_\mathrm{Ab}$\,=\,17.07mag. We do not assign any uncertainties to our magnitudes, since there was no direct photometric measurement and we just want to give a first prediction of possible photometric values. In addition the results are in a very good agreement to the absolute magnitude-spectral type relations of \citet{Liu06}.


\subsection{Testing evolutionary tracks}
\subsubsection{Ages}
\label{CMD}
%
  \begin{figure}
  \centering
       \includegraphics[width=8.8cm]{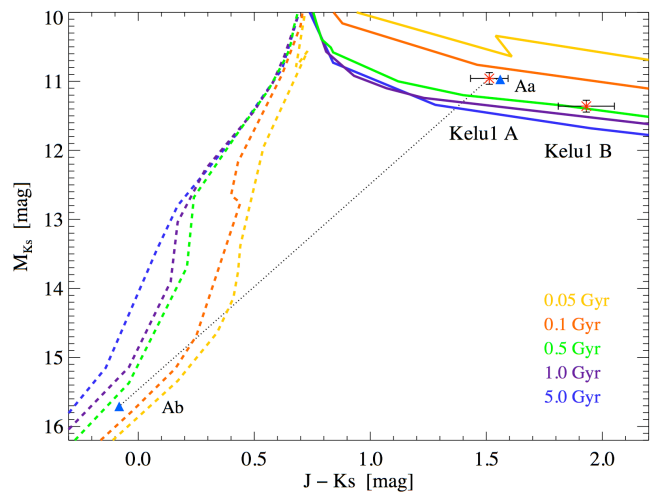}
  \caption{Color -- Magnitude diagram: The red asterisks show the location of the resolved components Kelu-1\,A and Kelu-1\,B. The position of the components Kelu-1\,Aa and Ab, if they were resolved, are indicated by the dotted lines and the triangles. While the DUSTY isochrones are displayed as solid lines in the upper right part, the COND isochrones are displayed by the dashed lines. }
   \label{age}
  \end{figure}
%
Age estimates of field brown dwarfs without established cluster membership have to rely on theoretical isochrones, using the mass-luminosity-age relation for brown dwarfs. Figure \ref{age} displays the color-magnitude diagram of Kelu-1\,A and B compared to the widely used DUSTY models of \citet{Chabrier00} for early to mid-type L dwarfs. For Kelu-1\,A these models predict an age clearly between 0.1 and 0.5 Gyr taking into account the estimated photometric errors. For Kelu-1\,B, however, they predict a slightly older age of 0.3 - 1 Gyr. Since binary components are believed to be coeval, we assess a minimum system age of 0.3 and a new upper age limit of 0.5 Gyr. In addition, the triangles in Fig. \ref{age} show the position of of the Kelu-1\,Aa and Kelu-1\,Ab components if they would be resolved. Therefore we included the COND models of \citet{Baraffe03} which better consider the dust condensation of T dwarfs. The resulting age for both individual components is consistent with the above estimation. 
The here derived age estimate (0.3-0.5 Gyr) further decreases the upper age limit for Kelu-1\,AB compared to previous estimates from \citet{Basri98_1} (0.3-1Gyr) and \citet{Liu} (03.-0.8 Gyr). The determination by \citet{Basri98_1} was solely based on the  strength of the unresolved \ion{Li}{i} absorption feature, while \citet{Liu} relied on the mass evolutionary tracks of the same model. They took the presence of lithium and the substellar nature of both components into account, which restricts the maximum mass for each component.

\subsubsection{Masses}
%
\begin{figure}
  \centering
       \includegraphics[width=8.8cm]{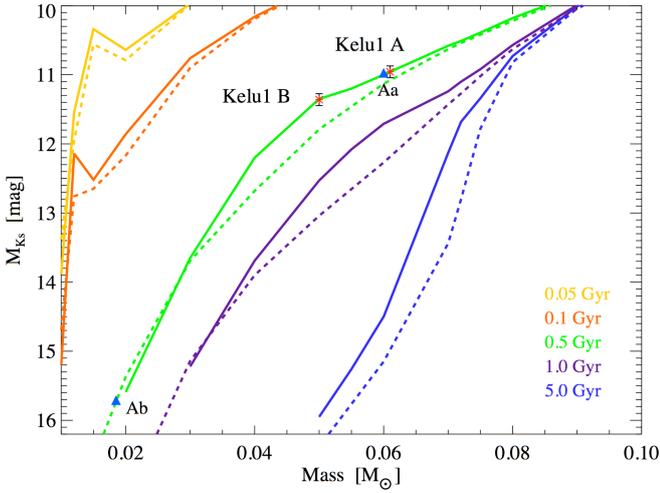}
  \caption{The absolute K$_\mathrm{S}$ magnitudes of the Kelu-1 system components compared to the DUSTY (solid lines) and COND (dashed lines) models at 0.05, 0.1, 0.5, 1 and 5 Gyr to derive maximum mass estimations for each component. }
   \label{CM_age}
  \end{figure}
%
In order to compare our dynamical mass determination of Kelu-1\,AB with mass predictions from evolutionary models, we again utilize the DUSTY and COND models as a function of absolute K$_\mathrm{S}$ magnitude and age (see Fig. \ref{CM_age}). Using the upper limit for the system age of 0.5 Gyr as obtained in \S\,\ref{CMD}, we derive maximum individual masses of 61\,$\pm$\,2\,$M_\mathrm{Jup}$ for Kelu-1\,A and 50\,$\pm$\,2\,$M_\mathrm{Jup}$ for Kelu-1\,B resulting in a maximum total system mass of 111\,$\pm$\,4\,$M_\mathrm{Jup}$. Even if we include the possible third component Kelu-1\,Ab, the additional mass of 18.5\,$M_\mathrm{Jup}$ would only increase the total mass to $\sim$\,130\,$M_\mathrm{Jup}$. This mass is still about 47\,$M_\mathrm{Jup}$ lower than the total system mass we derived in \S\,\ref{Orbit} by dynamical mass determination.
This discrepancy between luminosity, mass and age in the evolutionary tracks might result in an under-prediction of the masses.  A similar finding  has already been reported for other objects. Higher system masses were derived by dynamical mass determination, by \citet{Close05} for  AB Dor C, \citet{Ireland} for GJ 802 b or just recently \citet{Cardoso08} and Kasper et al. (ApJL, 2008 submitted) for  $\epsilon$\,Ind Ba, Bb. They all claim as possible reasons the underestimation of the abundance of condensates in the atmospheres as well as the effect of magnetic fields hindering heat flow or even the unusual assumption of non-coevality in binary systems.


\section{Summary and outlook}
We presented new astrometric and photometric observations of Kelu-1\,AB. For the first time we achieved resolved spectroscopy of the individual components of this brown dwarf binary system. The near-IR  spectra display interesting peculiarities for each component. The spectrum of Kelu-1\,A shows a distinct cup-shaped dip from 1.6 --1.65\,$\mu$m which did not fit any early, single L dwarf spectrum, but could be best reproduced by a combined spectrum of an early L- and late T dwarf. Our spectral template matching algorithm leads to the result that Kelu-1\,A itself is a close spectroscopic binary with spectral types of L0.5\,$\pm$\,0.5 and T7.5\,$\pm$\,1. In contrast, Kelu-1\,B is characterized by a triangular-shaped continuum, similar to very young red L dwarfs and was classified as L3\,pec $\pm$\,1.5\,.

Eight epochs of high-resolution astrometry from 2005-2008 enabled us to compute the orbital parameters and derive the first dynamical mass measurement for the system. We found a quite eccentric orbit ($e$ = 0.82\,$\pm$\,0.10 ), seen almost edge-on. The period of $38^{+8}_{-6}$ years and a semi-major axis of $6.4^{+2.4}_{-1.3}$ AU yields a total system mass of $177^{+113}_{-55}\,M_{\rm Jup}$ which is slightly higher than one would expect in a BD-BD binary system. If the system consists of only two objects, this would imply that the primary is more likely a very low mass star with a mass $\ga$ 85 $M_\mathrm{Jup}$. This is in contradiction to its defined early L-spectral type from spectroscopy which, along with its age estimation, indicates an affiliation to brown dwarfs. Including a possible third component in the system would solve this discrepancy with all component masses being below the substellar mass limit of 75 $M_\mathrm{Jup}$. The non-detection of this component in any previous high-resolution imaging observation (Keck, HST, NACO), indicates a maximum separation of $\sim$\,0.22$\arcsec$\,($\approx$\,4.1 AU) related to the assumed brightness difference of Kelu-1\,Aa and Ab. 

The derived magnitudes of L$\arcmin$\,=\,11.42\,$\pm$\,0.15 mag for Kelu-1\,A and L$\arcmin$\,=\,11.66\,$\pm$\,0.15 mag  for Kelu-1\,B provide the first resolved L-band photometry for any L dwarf binary. These values and the related near-IR colors support the spectral type determinations from spectroscopy. Comparing the derived near-IR photometry of the individual components with the widely used theoretical DUSTY models, led to a slightly younger age estimation of 0.3 -- 0.5 Gyr compared to previous estimations of its age. Using this new upper age limit, the evolutionary tracks yield only a maximum total system mass of $\sim$\,130 $M_\mathrm{Jup}$, even if a third companion is considered. This arising discrepancy to the dynamically determined mass shows the importance of further mass determinations independent of theoretical models.

With all the above conclusions, the hypothesis that Kelu-1\,A itself is a binary system seems conclusive. If the third component in the Kelu-1\,AB system can be confirmed in further observations, this would be the first pure brown dwarf triple system detected so far. 

Future work includes resolved optical spectra of Kelu-1\,AB. This will ascertain in which component Li absorption is present and thus help to better establish the ages of the components. In addition, it will extend the individual spectral type determinations into the red optical. This might give further conclusions on mechanisms like low-gravity, high-metallicity and/or cloud condensation leading to the redder, triangular shaped near-IR spectra of Kelu-1\,B. 
The continued monitoring of Kelu-1\,AB will help to refine the orbital parameters and its dynamical mass. Further, upcoming spatially resolved high-resolution radial velocity observations will provide information on the spectroscopic orbit for the determination of individual masses and possibly resolve the third component. 
These results will help to improve and adjust the theory of brown dwarf evolution and improve the interpretation of the observations of these objects.

\begin{acknowledgements}
 M.B.Stumpf acknowledges support by the \emph{DLR Verbundforschung} project number 50 OR 0401. H. Bouy acknowledges the funding from the European Commission's Sixth Framework Program as a Marie Curie Outgoing International Fellow (MOIF-CT-2005-8389). 
This work is based on observations collected at the European Southern Observatory (Paranal, Chile), programs 077.C-0791A+B, 078.C-0229, 079.C-0161, 081.C-0340, 380.C-0770 and 381.C-0527, PI M.B.Stumpf  and on observations made with the NASA/ESA Hubble Space Telescope, obtained at the Space Telescope Science Institute (STScI) with program GO-10208, PI W.Brandner. The \emph{HST} data set of reference PSF stars and previous Kelu-1 data were obtained from the \emph{HST} data archive at STScI. STScI is operated by the Association of Universities for Research in Astronomy, Inc., under NASA contract NAS 5-26555. We made also use of archive data from the \emph{Spitzer} Space Telescope, which is operated by JPL/Caltech under contract with NASA.

This research has benefitted from the SpeX Prism Spectral Libraries, maintained by Adam Burgasser at http://www.browndwarfs.org/spexprism.
This publication makes use of data products from the Two Micron All Sky Survey, which is a joint project of the University of Massachusetts and the Infrared Processing and Analysis Center/California Institute of Technology, funded by the National Aeronautics and Space Administration and the National Science Foundation.
\end{acknowledgements}

\bibliographystyle{aa}
\bibliography{literature.bib}

\end{document}